\documentstyle[12pt,epsfig]{article}

\evensidemargin=\oddsidemargin
\def\thefootnote{\fnsymbol{footnote}}
\newcommand{\ptmis}{{ {\rm p} \hspace{-0.53 em} \raisebox{-0.27 ex} {/}_T }}
\def\mathrm#1{\mbox{\rm #1}}

\newcommand{\noi}{\noindent}
\def\21{$SU(2) \otimes U(1)$}
\def\sm{\hbox{standard model }}
\def\VEV#1{\left\langle #1\right\rangle}

\def\lsim{\raise0.3ex\hbox{$\;<$\kern-0.75em\raise-1.1ex\hbox{$\sim\;$}}}
\def\gsim{\raise0.3ex\hbox{$\;>$\kern-0.75em\raise-1.1ex\hbox{$\sim\;$}}}

\def\np#1#2#3{           {\it Nucl. Phys. }{\bf #1} (19#2) #3}
\def\pl#1#2#3{           {\it Phys. Lett. }{\bf #1} (19#2) #3}
\def\prl#1#2#3{          {\it Phys. Rev. Lett. }{\bf #1} (19#2) #3}
\def\pr#1#2#3{           {\it Phys. Rev. }{\bf #1} (19#2) #3}
\def\ppnp#1#2#3{           {\it Prog. Part. Nucl. Phys. }{\bf #1} (19#2) #3}
\def\prep#1#2#3{         {\it Phys. Rep. }{\bf #1} (19#2) #3}
\def\nps#1#2#3{          {\it Nucl. Phys. B (Proc. Suppl.) }
    {\bf #1} (19#2) #3}
\def\mpl#1#2#3{          {\it Mod. Phys. Lett. }{\bf #1} (19#2) #3}

\begin{document}
\begin{titlepage}
\pagestyle{empty}
\rightline{hep-ph/9601269}
\rightline{FTUV/95-68}
\rightline{IFIC/95-71}
\rightline{KA-TP-12-1995} 
\rightline{MADPH-96-925}

\noindent January 1996 \hfill Submitted to Nucl. Phys. B

\vskip 1.0cm

\begin{center}
{\bf SEARCHING FOR INVISIBLY DECAYING HIGGS BOSONS \\
AT LEP II}
\vskip 1.cm
{\bf F.\ de Campos$^{a,}$\footnote{E-mail: fernando@flamenco.ific.uv.es},
O.\ J.\ P.\ \'Eboli$^{b,}$\footnote{Permanent address: Instituto de 
F\'{\i}sica, Universidade de S\~ao Paulo,
C.P.\ 66318, CEP 05389-970 S\~ao Paulo, Brazil.
E-mail: eboli@phenom.physics.wisc.edu}, 
J.\ Rosiek$^{a,c,}$\footnote{On leave of absence from Institute for 
Theoretical Physics, University of Warsaw. E-mail: rosiek@fuw.edu.pl,
rosiek@itpaxp3.physik.uni-karlsruhe.de},
and J.\ W.\ F.\ Valle$^{a,}$\footnote{E-mail: valle@flamenco.ific.uv.es}}
\vskip 0.5cm
$^a$Instituto de F\'{\i}sica Corpuscular - C.S.I.C., Dept.\ de F\'{\i}sica 
Te\`orica,\\
 Universitat de Val\`encia 46100 Burjassot, Val\`encia, Spain\\
\vskip 0.5cm
$^b$Physics Department, University of Wisconsin, Madison, WI 53706, USA\\
\vskip 0.5cm
$^c$Institut f\"ur Theoretische Physik, Universit\"at Karlsruhe\\
Postfach 6980, 76128 Karlsruhe, Germany

\vskip 1.cm
\baselineskip 14pt
\begin{quote}

  We study the potential of LEP II to unravel the existence of
  invisibly decaying Higgs bosons, predicted in a wide class of
  models. We perform a model independent analysis, focusing our
  attention to the final state topologies exhibiting $b \bar{b}$ or
  $\ell^+ \ell^-$ ($\ell=\mu$ or $e$) pairs and missing energy.  We
  carefully evaluate the signals and backgrounds, choosing appropriate
  cuts to enhance the discovery limits.  Our results demonstrate that
  LEP II is capable of discovering such a Higgs boson for a wide range
  of masses and couplings.

\end{quote}
\end{center}
\end{titlepage}
\vfill
\noi
\def\thefootnote{\arabic{footnote}}
\setcounter{footnote}{0}
\setcounter{page}{1}
\pagestyle{plain}
\baselineskip 18pt

\section{Introduction}

The problem of mass generation constitutes one of main puzzles in
particle physics. It is believed that spontaneous breaking of the
gauge symmetry through the expectation value of a scalar \21 doublet
is the origin of the masses of the fermions as well as those of the
gauge bosons. The key implication for this scenario is the existence
of the Higgs boson \cite{HIGGS}, not yet found. The first round of
$e^+ e^-$ collision experiments at LEP have constrained the \sm Higgs
boson mass to $m_h \gsim 65$ GeV \cite{RichardVal95}.  The second
phase of LEP will probe the electroweak breaking sector in a new
energy region and this is very interesting both from the point of view
of the standard model (SM) as well as its extensions.

A large variety of well motivated extensions of the SM Higgs sector
are characterised by the spontaneous violation of a global $U(1)$
lepton number symmetry by an \21 singlet vacuum expectation value
$\VEV{\sigma}$ \cite{CMP}.  In general, these models contain
additional Higgs bosons, as well as a massless Goldstone boson, called
majoron ($J$), which interacts very weakly with normal matter, and
have been postulated in order to give mass to neutrinos in various
different contexts \cite{fae}.  It is specially interesting for our
purposes to consider those models where such symmetry is broken at the
electroweak scale or below, i.e. $\VEV{\sigma} \lsim {\rm O}(1)$ TeV
\cite{HJJ}.  Although the interactions of the majoron with quarks,
leptons, and gauge bosons is naturally very weak, as required by
astrophysics \cite{KIM}, it can have a relatively strong interaction
with the Higgs boson. In this case the main Higgs boson decay channel is
likely to be ``invisible'', e.g.
\begin{equation}
h \to J J \; ,
\end{equation}
where $J$ denotes the majoron field. This feature also appears in
variants of the minimal supersymmetric model in which $R$ parity is
broken spontaneously \cite{MASI_pot3}.
Notwithstanding, our discussion is not limited to majoron models
since invisibly decaying Higgs bosons also appear in other models
\cite{suzuki}. 
For instance, in the minimal supersymmetric standard model with
conserved $R$ parity, the Higgs boson can decay invisibly into the lightest
neutralino pair depending on the choice of the parameters.

The invisible Higgs boson decay leads to events with large missing
energy that could be observable at LEP II and affect the Higgs boson
discovery limits. In particular, the invisible decay could contribute
to the signal of two acoplanar jets or leptons plus missing momentum.
This feature of invisible Higgs boson models allows one to strongly
constrain the Higgs boson mass in spite of the fact that the model involves
new parameters compared to the ones of the SM.  In particular, the LEP
I limit on the predominantly doublet Higgs boson mass is close to the SM
limit irrespective of the decay mode of the Higgs boson
\cite{valle:1,dproy}.

In the next section, we discuss the parameterisation of Higgs bosons
couplings relevant for their production at LEP.  Section \ref{sigback}
contains a detailed presentation of the expected Higgs boson signals
as well as SM backgrounds in the simplest majoron extensions of the
standard model, which contain both the $Z h$ as well as $A h$
production channels.  Section 4 contains a discussion of the Higgs
boson discovery limits at LEP II for the various topologies considered
in section 3 and for different LEP II centre-of-mass energies. In the
last section we present a brief overall discussion of the
phenomenological implications.

\section{Parameterisation of Higgs Boson Production and Decays}

Here we consider a model containing two Higgs doublets ($\phi_{1,2}$)
and a singlet ($\sigma$) under the \21 group.  The singlet Higgs field
carries a non-vanishing global lepton number charge. The scalar Higgs
potential of the model can be specified as:
\begin{eqnarray}
\label{N2}
V &=&\mu_{i}^2\phi^{\dagger}_i\phi_i+\mu_{\sigma}^2\sigma^{\dagger}
\sigma + \lambda_{i}(\phi^{\dagger}_i\phi_i)^2+
   \lambda_{3} (\sigma^{\dagger}\sigma)^2+
\nonumber\\
& &\lambda_{12}(\phi^{\dagger}_1\phi_1)(\phi^{\dagger}_2\phi_2)
+\lambda_{13}(\phi^{\dagger}_1\phi_1)(\sigma^{\dagger}\sigma)+
\lambda_{23}(\phi^{\dagger}_2\phi_2)(\sigma^{\dagger}\sigma)
\nonumber\\
& & +\delta(\phi^{\dagger}_1\phi_2)(\phi^{\dagger}_2\phi_1)+
    \frac{1}{2}\kappa[(\phi^{\dagger}_1\phi_2)^2+\;\;h.\;c.]
\end{eqnarray}
where the sum over repeated indices $i$ (=1,2) is assumed.

For appropriate choice of parameters, the minimisation of the above
potential leads to the spontaneous breaking of the \21 gauge symmetry,
as well as the global $U(1)_L$ symmetry. This allows us to identify a
total of three massive CP even scalars $H_{i} \:$ (i=1,2,3), plus a
massive pseudoscalar $A$ and the massless majoron $J$%
\footnote{For simplicity we assume throughout this 
paper that CP is conserved in the scalar sector.}.
For definiteness we assume that at the LEP II energies only three
Higgs particles can be produced: the lightest CP-even scalar $h$, the
CP-odd massive scalar $A$, and the massless pseudoscalar majoron $J$.
Notwithstanding, our analysis is also valid for the situation where
the Higgs boson $A$ is absent \cite{valle:2}, which can be obtained by
setting the couplings of this field to zero.

At LEP II, the main production mechanisms of invisible Higgs bosons
are the Bjorken process ($e^+e^- \rightarrow h Z$) and the associated
production of Higgs bosons pairs ($e^+e^- \rightarrow A h$), which
rely upon the couplings $hZZ$ and $hAZ$ respectively. An important
feature of the above model is that the majoron is a singlet under
\21 and possesses feable couplings to the gauge bosons, 
thus evading strong LEP I constraints coming from the 
invisible $Z$ width. The $hZZ$ and $hAZ$ interactions can be 
expressed, without loss of generality, in terms of the two 
parameters $\epsilon_A$ and $\epsilon_B$:
\begin{eqnarray}
\label{HZZ3}
{\cal L}_{hZZ}
&=& \epsilon_B \left ( \sqrt{2}~ G_F \right )^{1/2}
M_Z^2 Z_\mu Z^\mu h 
\; , 
\\
{\cal L}_{hAZ}&=& - \epsilon_A \frac{g}{\cos\theta_W} 
Z^\mu h \stackrel{\leftrightarrow}{\partial_\mu} A 
\; ,
\end{eqnarray}
with $\epsilon_{A(B)}$ being determined once a model is chosen.  For
instance, in the framework of the minimal SM $\epsilon_A=0$ and
$\epsilon_B=1$, while a majoron model with one doublet and one singlet
leads to $\epsilon_A=0$ and $\epsilon_B^2 \leq 1$. In the framework of
the minimal supersymmetric standard model  $\epsilon_{A(B)}$ 
are functions of the parameters defining this model.

The signatures of the Bjorken process and the associated production
depend upon the allowed decay modes of the Higgs bosons h and A. For
Higgs boson masses $m_h$ accessible at LEP II energies the main decay
modes for the CP-even state $h$ are $b \bar{b}$ and $JJ$. We treat the
branching fraction $B$ for $h \rightarrow JJ$ as a free parameter.  In
most models $B$ is basically unconstrained and can vary from 0 to 1.
Moreover, we also assume that, as it happens in the simplest models,
the branching fraction for $A \rightarrow b \bar{b}$ is nearly one,
and the invisible $A$ decay modes $A\rightarrow hJ$, $A\rightarrow
JJJ$, although CP-allowed, do not exist. Therefore our analysis
depends finally upon five parameters: $M_h$, $M_A$, $\epsilon_A$,
$\epsilon_B$, and $B$. This parameterisation is quite general and very
useful from the experimental point of view since limits on $M_h$,
$M_A$, $\epsilon_A$, $\epsilon_B$, and $B$ can be later translated
into bounds on the parameter space of many specific models.

The parameters defining the general $hZZ$ and $hAZ$ interactions can
be constrained by the LEP I data. In fact, Refs.\ 
\cite{valle:1,valle:asso} analyse some signals for invisibly decaying
Higgs bosons, and conclude that LEP I excludes $M_h$ up to 60 GeV
provided that $\epsilon_B > 0.4$.  In what follows we extend the
analysis to the energies that will be available at LEP II.

%^^^^^^^^^^^^^^^^^^^^^^^^^^^^^^^^^^^^^^^^^^^^^^^^^^^^^^^^^^^^^^^^^^^^^

\section{Signatures and backgrounds}
\label{sigback}

\begin{figure}[p]
\begin{center}
\mbox{\epsfig{file=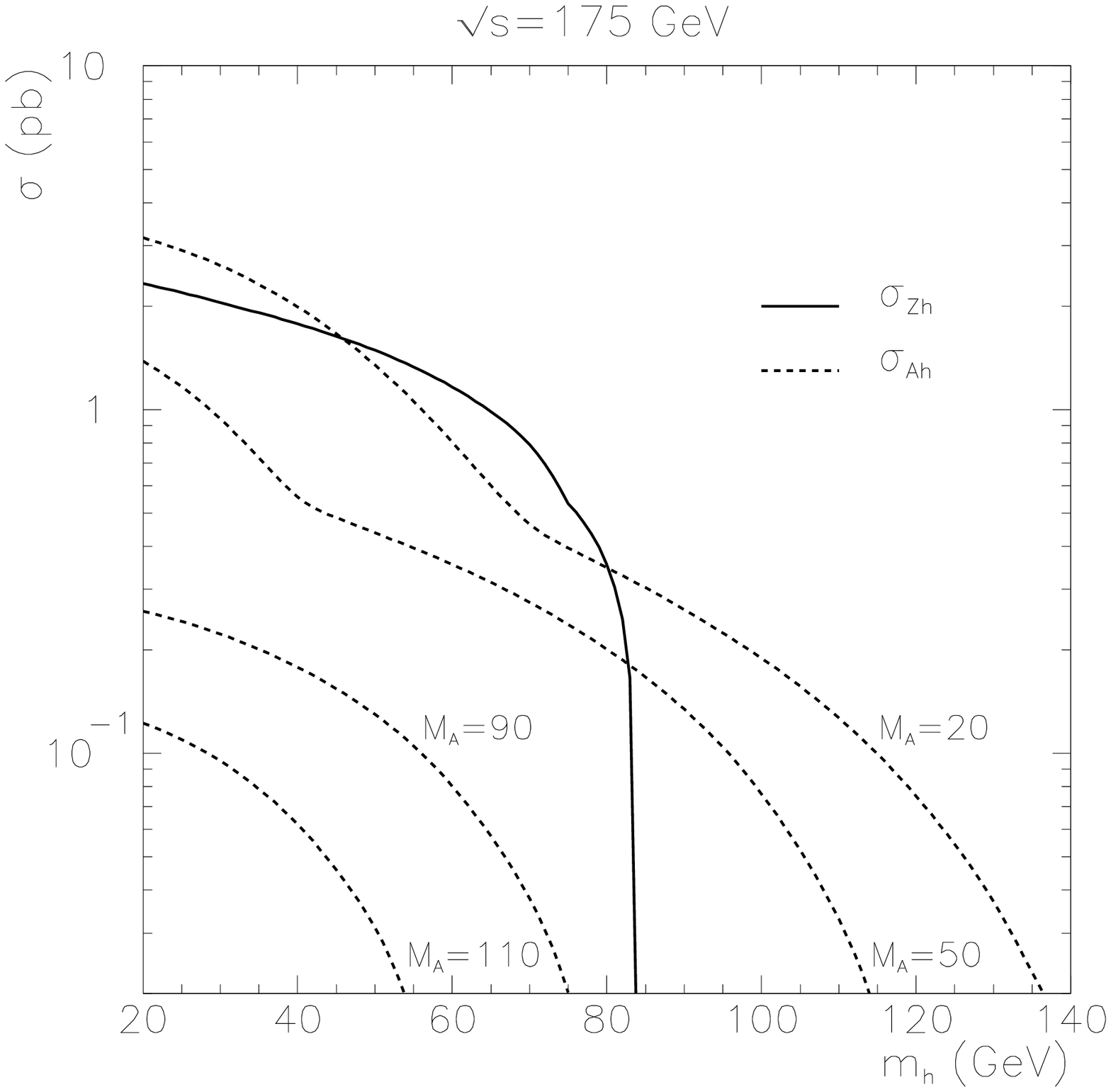,width=0.6\linewidth}}
\caption{\protect\small
\baselineskip 12pt 
Total cross section for the production of invisibly decaying Higgs
bosons through the Bjorken (solid line) and associated production
(dotted line) mechanisms at $\protect\sqrt{s}=175$ GeV.}
\label{fig:sigma}
\end{center}
\end{figure}

In this work, we focus our analysis in the following signals for the
production of invisibly decaying Higgs bosons h:
\begin{eqnarray}
e^+ e^-  &\rightarrow & (Z h + A h) \rightarrow b \bar{b}~+~ \ptmis
\; ,
\label{bb:inv} 
\\
e^+ e^-  &\rightarrow & Z h \rightarrow \ell^+ \ell^- ~+~ \ptmis
\; ,
\label{ll:inv}
\end{eqnarray}
where $\ell$ stands for $e$ or $\mu$. The signal (\ref{ll:inv}) was
previously analysed in Refs.\ \cite{valle:1} and \cite{valle:2}.  At
LEP II energies the $W$--fusion process ($e^+ e^- \rightarrow \nu_e
\bar\nu_e h$) leads not only to a negligible contribution to the Higgs
production cross section but also to an unidentifiable final state,
since $h \rightarrow JJ$, and consequently we will not take this
reaction into account.  We exhibit in Fig.\ \ref{fig:sigma} the total
cross section for the production of $Zh$ and $Ah$ pairs before the
introduction of cuts, assuming that $\epsilon_A = \epsilon_B = 1$. It
is interesting to notice that the associated production dominates over
the Bjorken mechanism for $M_A < M_Z$. This effect is further
enhanced by the large branching fraction of $A$ going into $b$-quark
pairs and of $h$ going to $JJ$.

For the sake of completeness,  we also include the channels
where $h$ decays visibly into a $b\bar{b}$ quark pair
\begin{eqnarray}
e^+ e^-  &\rightarrow & Z h \rightarrow \ell^+ \ell^-  +~b \bar{b} 
\; ,
\label{ll:bb} 
\\
e^+ e^-  &\rightarrow & (Z h + A h) \rightarrow b \bar{b}~ +~b \bar{b}
\; ,
\label{bb:bb}
\end{eqnarray}
which allow us to obtain additional limits on $\epsilon_A$ and
$\epsilon_B$.  These channels were subject of many detailed analyses
performed in the framework of the SM or the two Higgs doublet
model. Thus, we do not repeat them fully here.  Instead, we adopt
partially the results quoted in Ref.\ \cite{DELPHI} and combine them with
our results on the invisible Higgs decay channels.

Our goal is to evaluate the limits on $M_h$, $M_A$, $\epsilon_A$,
$\epsilon_B$, and $B$ that can be obtained at LEP II from the above
processes.  In order to do so, we study carefully the signals and
backgrounds, choosing the cuts to enhance the former. We analyse the
signals and backgrounds using the PYTHIA event generator
\cite{PYTHIA}, and taking into account the QED (QCD) initial and final
state radiation, as well as fragmentation. In order to reconstruct the
jets we employ the subroutine LUCLUS of PYTHIA.

%^^^^^^^^^^^^^^^^^^^^

\subsection{ Topology $b\bar{b}\protect\ptmis$ }
\label{subsec:bbmis}

There are three sources of signal events with the topology 2 $b$-jets
+ $\ptmis$: one due to the associated production of Higgs bosons and
two due to the Bjorken mechanism.
\begin{eqnarray}
e^+  e^- &\rightarrow &( A  \rightarrow b   \bar{b} )~   
+ ~(h \rightarrow  J  J) \; ,
\label{h:a}
\\
e^+  e^- &\rightarrow& ( Z  \rightarrow b   \bar{b} )~   
+ ~(h \rightarrow  J  J) \; ,
\label{h:jj}
\\
e^+  e^- &\rightarrow& ( Z  \rightarrow \nu \bar{\nu} )~ 
+ ~(h \rightarrow  b \bar{b}) \; .
\label{h:sm}
\end{eqnarray}

In the framework of the SM, there are several sources of background
for this topology%
\footnote{We did not take into account the
non-resonant contributions to the process $e^+ e^- \rightarrow Z~\nu
\bar{\nu}\rightarrow q \bar{q}~\nu \bar{\nu}$ since they are small at
LEP II energies~\cite{mele} compared with the resonant process
(\ref{zz:bbinv}).}:
\begin{eqnarray}
e^+ e^-  &\rightarrow& Z/\gamma~ Z/\gamma \rightarrow q \bar{q} 
~\nu \bar{\nu}
\; ,
\label{zz:bbinv} 
\\
e^+ e^-  &\rightarrow& Z^*/\gamma^*~  \rightarrow q \bar{q} 
~[n \gamma ]
\; ,
\label{z:bbinv} 
\\
e^+ e^-  &\rightarrow&  [e] \gamma e \rightarrow [e] \nu W
\rightarrow q \bar{q}^\prime 
~ [e] \nu
\; ,
\label{ewn:bbinv} 
\\
e^+ e^-  &\rightarrow& W^+ W^- \rightarrow q \bar{q}^\prime ~ [\ell] \nu
\; ,
\label{ww:bbinv} 
\\
e^+ e^-  &\rightarrow& [e^+e^-] \gamma\gamma \rightarrow [e^+e^-]~ q \bar{q}
\; ,
\label{gg:bbinv} 
\end{eqnarray}
where the particles in square brackets escape undetected and the jet
originating from the quark $q$ is identified (misidentified) as being
a $b$-jet.  In our analysis, we assume that particles making an angle
smaller than 12$^\circ$ with the beam pipe are not detected. The above
reactions exhibit two sources of missing momentum: neutrinos and
particles going down the beam pipe. Moreover, the final state jets can
also lead to missing transverse momentum since we perform a full
simulation of the event, allowing for meson and hadron decays that can
produce neutrinos or undetected particles. The expected numbers of
background events from the processes (\ref{zz:bbinv} --
\ref{gg:bbinv}), before applying the selection cuts, are shown in the
Table~\ref{tab:bbnc}.

\begin{table}[htbp]
\begin{center}
\begin{tabular} {|c|c|c|c|c|c|c|}
\hline
$\sqrt{s}$ (GeV)& ${\cal L}$ (pb$^{-1}$)& $Z/\gamma~Z/\gamma$
& $Z^*/\gamma^*$ & $e\nu W$& $W^+W^-$ & $\gamma\gamma$
\\
\hline
175 & 500 & 105 & 5.5~10$^4$ & 220 & 6.4~10$^3$ & 2.2~10$^3$\\
190 & 300 & 209 & 2.6~10$^4$ & 182 & 4.9~10$^3$ & 1.4~10$^3$  \\
205 & 300 & 295 & 2.2~10$^4$ & 237 & 5.1~10$^3$ & 1.5~10$^3$    \\
\hline
\end{tabular}
\caption{ \protect \small \baselineskip 12pt
Expected number of background events in the
$b\bar{b} \protect\ptmis$ channel before cuts for three values of
$\protect\sqrt{s}$ and integrated luminosity ${\cal L}$.}
\label{tab:bbnc}
\end{center}
\end{table}

At this point the simplest and most efficient way to improve the
signal-to-background ratio is to use the fact that the Higgs bosons
$A$ and $h$ decay into jets possessing $b$-quarks. So we require that
the events contain $b$-tagged jets. Moreover, the background can be
further reduced by demanding a large $\ptmis$. Having these facts in
mind we impose the following set of cuts, based on the ones used by
the DELPHI collaboration for the SM Higgs boson search~\cite{DELPHI}:

\begin{enumerate}

\item {\em Missing momentum cuts.} We require:

  \begin{itemize}

  \item The $z$ component of the missing momentum to be smaller than
    $0.15\times \sqrt{s}$.

  \item The absolute value of the cosine of the polar angle of the
    missing momentum to be less than 0.9. These two cuts are used to
    reject events whose missing momentum is due to undetected
    particles going down the beam pipe.

  \item The transversal component of missing momentum $\ptmis$ should be
    bigger than 25 GeV for $\sqrt{s}=175$ and $190$ GeV and 30 GeV for
    $\sqrt{s}=205$ GeV.

  \end{itemize}

\item {\em Acolinearity cut.} The cosine of the angle between the axes
  of the two most energetic jets is required to be above -0.8. This is
  equivalent to the requirement that the angle between the jet axes is
  smaller than $145^{\circ}$. This cut reduces the $Z \rightarrow q
  \bar{q}$ background, where the $\ptmis$ originates from neutrinos
  and jet fluctuations, and consequently it is parallel to the jet
  thrust axes.

\item {\em Scaled acoplanarity cut.} Acoplanarity is defined as the
  complement of the angle in the plane perpendicular to the beam
  between the total momenta in the two thrust hemispheres. Scaling the
  acoplanarity by the minimum of $\sin\theta_{jet~1}$ and $
  \sin\theta_{jet~2}$ \cite{DELPHI} avoids instabilities at low polar
  jet angles. We impose that the scaled acoplanarity is greater than
  $7^{\circ}$.

\item {\em Thrust/number of jets cut.} We require the event thrust to be
  bigger than 0.8. However, this cut gives relatively small signal
  efficiency for the process (\ref{h:a}) (or (\ref{h:sm})) provided
  $M_A$ ($M_h$) is in the range $45-80$ GeV. Therefore, for this mass
  range, we demand that the two most energetic jets should carry more
  than 85\% of the visible energy instead of the thrust cut.

\item {\em Charged multiplicity cut.} We impose that the event should
  contain more than 8 charged particles. This cut eliminates potential
  backgrounds from the production of $\tau^+ \tau^-$ pairs.

\item {\em $b$-tagging cut.} We accept only the events containing 2
  $b$-tagged jets. In the analysis, we adopt the efficiencies for the
  $b$-tagging directly from the DELPHI note \cite{DELPHI}: 68\%
  efficiency for the signal and the appropriate values for the
  backgrounds extracted from Table~5 of Ref.\ \cite{DELPHI}.

\item {\em Invariant mass cut.} We impose that the total visible
  invariant mass should be in the range $M \pm 10$ GeV, where $M$ is
  the mass of the visibly decaying particle ($Z$, $h$ or $A$).

\end{enumerate}

\begin{figure}[htbp]
\begin{center}
\mbox{\epsfig{file=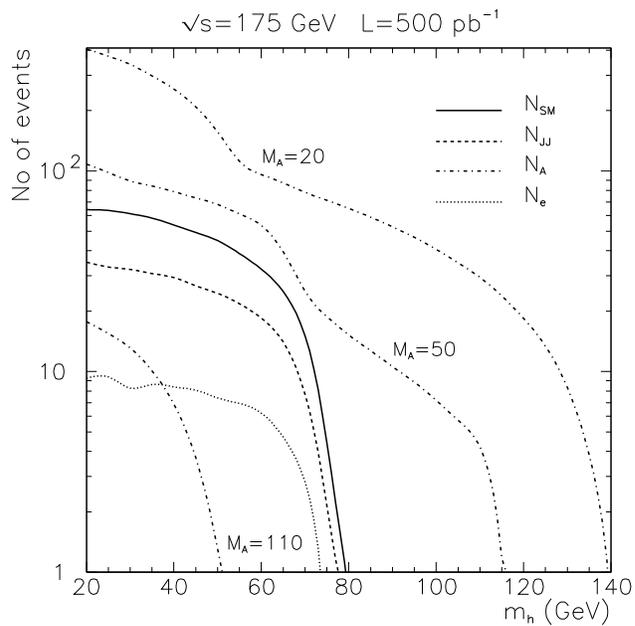,width=0.6\linewidth}}
\caption{\protect\small
\baselineskip 12pt
Expected number of signal events after cuts for the processes
$e^+  e^- \rightarrow ( Z  \rightarrow b   \bar{b} )   
+ (h \rightarrow  J  J)$ ($N_{JJ}$),
$e^+  e^- \rightarrow ( Z  \rightarrow \nu \bar{\nu} ) 
+ (h \rightarrow  b \bar{b})$ ($N_{SM}$),
$e^+  e^- \rightarrow ( A  \rightarrow b   \bar{b} )   
+ (h \rightarrow  J  J)$  ($N_A$), and
$e^+  e^- \rightarrow ( Z  \rightarrow e^+ e^- )
+ (h \rightarrow  b \bar{b})$ ($N_e$),
assuming $\epsilon_A=\epsilon_B=1$ and no suppression due to $h$ decay
branching ratios in each case. Note that $N_A$ is given for three
choices of $M_A$ values.}
\label{fig:sig}
\end{center}
\end{figure}

We exhibit in Fig.\ \ref{fig:sig} the expected number of signal events
$N_A$, $N_{JJ}$, and $N_{SM}$ originating from the production
processes (\ref{h:a}), (\ref{h:jj}), and (\ref{h:sm}) respectively,
for $\sqrt{s}=175$ GeV and an integrated luminosity ${\cal L}=500$
pb$^{-1}$.  We impose all the above cuts, but the invariant mass one,
and assume that $\epsilon_A = \epsilon_B = 1$ and that there is no
suppression due to the $h$ decay branching ratio ($B$).  Obviously, it
is trivial to obtain the number of signal events for arbitrary
$\epsilon_A$, $\epsilon_B$, and $B$ from this figure by re-scaling 
our results with appropriate powers of these parameters; see
Sec.\ 3 below.

The number of background events after applying the above cuts, excluding
the invariant mass cut, are shown in Table~\ref{tab:bbc}. The most
important background after the cuts is the production of a $Z$ pair
(\ref{zz:bbinv}), which grows substantially after the threshold for
the production of on-shell $Z$'s is reached.  Notice that our cuts
eliminate completely the large background due to $\gamma\gamma$
reactions.

\begin{table}[htbp]
\begin{center}
\begin{tabular}{|c|c|c|c|c|c|c|c|}
\hline
$\sqrt{s}$ (GeV) & ${\cal L}$ (pb$^{-1}$) & $Z/\gamma~Z/\gamma$
& $Z^*/\gamma^*$ & $e \nu W$ & $W^+W^-$ & $\gamma\gamma$ &  Total \\
\hline
175 & 500 & 0.79 & 0.76 & 0.46 & 0.29 & 0.00 & 2.31 \\
190 & 300 & 1.17 & 0.44 & 0.38 & 0.23 & 0.00 & 2.23 \\
205 & 300 & 4.26 & 0.12 & 0.46 & 0.19 & 0.00 & 5.02 \\
\hline
\end{tabular}
\caption{  \protect \small \baselineskip 12pt
Number of the background events in the $b\bar{b}\protect\ptmis$ 
channel after all cuts, but the invariant mass one.}
\label{tab:bbc}
\end{center}
\end{table}

The backgrounds can be further reduced introducing the visible
invariant mass cut. However, depending on the $h$ and $A$ masses, this
cut also reduces the signal and weakens the limits on the $ZhA$ and
$ZZh$ couplings. Therefore, for each mass combination four limits are
calculated: with or without the invariant mass cut and with the thrust
cut or the cut on the minimal two-jet energy. The best limit is kept.

%^^^^^^^^^^^^^^^^^^^^

\subsection{ Topology $\ell^+\ell^-\protect\ptmis$ }

The events with the final state topology $\ell^+ \ell^- \ptmis$ are
generated by the Bjorken process
\begin{eqnarray}
e^+  e^- &\rightarrow& ( Z  \rightarrow \ell^+ \ell^- )~   
+ ~(h \rightarrow  J  J)  \; ,
\label{l:jj}
\end{eqnarray}
where $\ell = e$ or $\mu$.  In this case, the signature is the
presence of two charged leptons with an invariant mass compatible with
the $Z$ mass, plus missing energy. This topology is the trademark of
all models exhibiting invisibly decaying Higgs bosons with sizable
couplings to the $Z$. Notice that the cross section for this process
depends only upon $\epsilon_B$, $M_h$, and $B$.

We consider only the $e^+e^-$ and $\mu^+\mu^-$ channels because they
are cleaner than the $\tau^+ \tau^-$ one and their backgrounds are
smaller. The possible background sources for this topology are
\begin{eqnarray}
e^+ e^-  &\rightarrow& Z/\gamma~ Z/\gamma ~\rightarrow  ~ 
\ell^+ \ell^- ~ \nu \bar{\nu}
\; ,
\label{zz:llinv} 
\\
e^+ e^-  &\rightarrow& Z^*/\gamma^*~  \rightarrow 
~\ell^+ \ell^- ~[n \gamma ]
\; ,
\label{z:llinv} 
\\
e^+ e^-  &\rightarrow& W^+ W^- ~\rightarrow~
 \ell^+ \ell^-  ~ \nu \bar{\nu}
\; ,
\label{ww:llinv} 
\\
e^+ e^-  &\rightarrow& e \gamma e ~\rightarrow~ 
e W \nu  ~\rightarrow~ e^\pm \ell^\mp ~\nu \bar{\nu}
\; ,
\label{ewnu:elinv}
\\
e^+ e^-  &\rightarrow& [e^+e^-]~ \gamma\gamma \rightarrow ~
[e^+e^-]~ \ell^+ \ell^- 
\label{gg:llinv} 
\; .
\end{eqnarray}
Notice that the background (\ref{ewnu:elinv}) is relevant just for
$\ell=e$.

The $\ell^+ \ell^- \ptmis$ signal shares many features in common with
the $b \bar{b} \ptmis$ one. First of all, the presence of an invisibly
decaying particle leads to missing energy. Furthermore, the two
leptons in the final state are not collinear nor in the same plane
since this is not a two going into two process. Therefore, in order to
enhance the signal, we introduce the following cuts similar to the
ones applied for the $b \bar{b} \ptmis$ topology:

\begin{enumerate}

\item We require the events to satisfy the same missing momentum cuts
  employed in Sec.\ 3.1.

\item We introduce an acolinearity cut, imposing that the 
cosine of the angle between the leptons is larger than -0.8. 

\item We also demand the scaled acoplanarity of the lepton pair to be
  greater than $7^\circ$.

\item We require that the event should contain exactly 2 charged
  particles identified as electrons or muons.

\item We impose that the invariant mass of the lepton pair
should be in the range $M_Z \pm 5$ GeV. In addition, we require that the
total energy of the lepton pair should be in the range $E_Z(M_h) \pm 5$ GeV,
where
\begin{eqnarray}
E_Z(M_h) = {s + M_Z^2 - M_h^2 \over 2 \sqrt{s}} \; .
\label{ll:zen} 
\end{eqnarray}
These cuts are essential to reduce the $WW$ background.  Notice that
the invariant mass bin for the lepton pair in this topology is smaller
than the one used for the visible mass for the topology $b\bar{b}
\ptmis$ because the energy and momentum can be better
determined for leptons than for jets.

\end{enumerate}

The expected number of $e^+e^- \ptmis $ signal events ($N_e$) after
cuts is also shown for $\sqrt{s}=175$ GeV in Fig.\ \ref{fig:sig}.
Notice that for a wide range of masses $M_h$ and $M_A$ there are more
signal events with the topology $b\bar{b}\ptmis$. We exhibit in
Table~\ref{tab:ee} the expected number of background events
originating from the processes (\ref{zz:llinv}--\ref{ewnu:elinv}),
before and after applying the above cuts.  Notice that the most
important irreducible background after the cuts is due to process
(\ref{ww:llinv}). Two photon reactions, process (\ref{gg:llinv}), lead
to a large number of $\ell^+\ell^-$ pairs (3200, 2080, and 2230 at
$\sqrt{s}=$ 175, 190, and 205 GeV respectively), however, it is
completely eliminated by our cuts.

\begin{table}[htbp]
\begin{center}
\begin{tabular} {|c|c|c|c|c|c|c|c|c|c|c|}
\hline
 & & 
\multicolumn{2}{|c|}{$Z/\gamma~Z/\gamma$} & 
\multicolumn{2}{|c|}{$Z^*/\gamma^*$} & 
\multicolumn{2}{|c|}{$W^+W^-$ } & 
\multicolumn{2}{|c|}{$e\nu W$ } & 
Total \\
\cline{3-11}
$\sqrt{s}$ (GeV) & ${\cal L}$ (pb$^{-1}$) 
& (a) & (b) & (a) & (b) & (a) & (b) & (a) & (b) & (b) \\
\hline
175 & 500 & 6.4 & 0.01 & 4.2~10$^3$ & 0.00 & 74 & 2.47 & 232 & 0.01 & 2.49 \\
190 & 300 & 14  & 0.14 & 2.1~10$^3$ & 0.00 & 57 & 1.03 & 189 & 0.02 & 1.19 \\
210 & 300 & 20  & 1.25 & 1.7~10$^3$ & 0.00 & 60 & 0.73 & 242 & 0.01 & 2.00 \\
\hline
\end{tabular}
\caption{  \protect \small \baselineskip 12pt
Numbers of the background events in the $e^+e^- \protect\ptmis$
channel. (a) and (b) denote the number of events before and after cuts
respectively. Since the number of events after cuts depend on $M_h$,
we display the maximal values. The backgrounds for the $ \mu^+\mu^-
\protect\ptmis$ channel are identical except for the $e\nu W$ column, which
vanishes in this case.}
\label{tab:ee}
\end{center}
\end{table}

%^^^^^^^^^^^^^^^^^^^

\subsection{Topologies without missing energy}

There are three signal processes where the amount of missing energy
should be small, up to initial state radiation and jet
fluctuations. They are
\begin{eqnarray}
e^+  e^- &\rightarrow& ( A  \rightarrow b   \bar{b} )~   
+ ~(h \rightarrow b   \bar{b} ) \; ,
\label{a:bb}
\\
e^+  e^- &\rightarrow& ( Z  \rightarrow b \bar{b} )~   
+ ~(h \rightarrow b \bar{b} ) \; ,
\label{z:bb}
\\
e^+  e^- &\rightarrow& ( Z  \rightarrow \ell^+ \ell^- )~
+ ~(h \rightarrow  b \bar{b}) \; .
\label{z:ll}
\end{eqnarray}
Therefore, we must consider two new topologies: events with 4
$b$-tagged jets ($b\bar{b}b\bar{b}$) and events exhibiting 2 leptons
and 2 $b$-jets ($\ell^+\ell^-b\bar{b}$). These topologies were the
subject of many extensive analyses within the framework of the SM and
its minimal supersymmetric version \cite{DELPHI}.

For the sake of completeness, we take into account the signal events
originating from processes (\ref{a:bb}--\ref{z:ll}). However, we only
evaluate the total signal cross sections without cuts using the PYTHIA
7.4 generator. In order to study the constraints emanating from these
processes we adopt the signal detection efficiencies and the estimated
background values quoted in Ref.\ \cite{DELPHI}.

It is interesting to point out that the ratio of the number of events
with topology $b\bar{b}b\bar{b}$ to the ones with $\ell^+ \ell^- b
\bar{b}$ is independent of $B$.

%^^^^^^^^^^^^^^^^^^^^^^^^^^^^^^^^^^^^^^^^^^^^^^^^^^^^^^^^^^^^^^^^^^^^^

\section{Bounds on invisibly decaying Higgs boson couplings}

We define in Table~ \ref{tab:labels} the symbols used to denote the
number of signal events for the different topologies analysed in the
previous section, after imposing the cuts and assuming that
$\epsilon_A = \epsilon_B = 1$ and that there is no suppression due to
the $h$ branching ratio to each final state.

\begin{table}[htbp]
\begin{center}
\begin{tabular}{|l|l|}
\hline
Symbol & Process \\
\hline
$N_{JJ}(M_h)$ & $e^+e^-\rightarrow (Z\rightarrow b\bar{b}) 
+ (h\rightarrow JJ)$ \\ 
%\hline
$N_{SM}(M_h)$ & $e^+e^-\rightarrow (Z\rightarrow\nu\bar{\nu}) 
+ (h\rightarrow b\bar{b})$ \\ 
%\hline
$N_L(M_h)$ & $e^+e^-\rightarrow (Z\rightarrow \ell^+\ell^-)
+ (h\rightarrow  JJ)$ \\ 
%\hline
$N_{ZH}(M_h)$ & $e^+e^-\rightarrow (Z\rightarrow b\bar{b}) 
+ (h\rightarrow b\bar{b})$ \\ 
%\hline
$N_{ZL}(M_h)$ & $e^+e^-\rightarrow (Z\rightarrow \ell^+\ell^-) 
+ (h\rightarrow b\bar{b})$ \\ 
%\hline
$N_A(M_h,M_A)$ & $e^+e^-\rightarrow (A\rightarrow b\bar{b})
+ (h\rightarrow JJ)$ \\ 
%\hline
$N_{AH}(M_h,M_A)$ & $e^+e^-\rightarrow (A\rightarrow b\bar{b}) 
+ (h\rightarrow b\bar{b})$ \\ 
\hline
\end{tabular}
\caption{ \protect \small \baselineskip 12pt
Symbols used to denote the number of signal events after cuts in the
various channels.}
\label{tab:labels}
\end{center}
\end{table}

The expected numbers of signal events for the various final state
topologies can be expressed as simple combinations of the parameters 
$\epsilon_A$, $\epsilon_B$, and $B$ and the quantities defined in 
Table~ \ref{tab:labels}, which, in turn, depend on the Higgs boson
masses $(M_h,M_A)$:
\begin{eqnarray}
&N_{bb}(M_h,M_A) &=~ \epsilon_B^2 \left [ B N_{JJ} + (1-B) N_{SM} \right ]
+ \epsilon_A^2 B N_A 
\; ,
\label{eq:nbb}\\
&N_{ll}(M_h) &=~ \epsilon_B^2 B N_L 
\; ,
\label{eq:nll}\\
&N_{4b}(M_h,M_A) &=~ \epsilon_B^2 (1-B) N_{ZH} + \epsilon_A^2 (1-B) N_{AH} 
\; ,
\label{eq:n4b}\\
&N_{llbb}(M_h) &=~ \epsilon_B^2 (1-B) N_{ZL}
\; ,
\label{eq:nllbb}
\end{eqnarray} 
where the quantities $N_{bb}$, $N_{\ell\ell}$, $N_{4b}$, and
$N_{llbb}$ stand for the number of signal events after cuts for the
topologies $b \bar{b} \ptmis$, $\ell^+ \ell^- \ptmis$, $b \bar{b} b
\bar{b}$, and $\ell^+ \ell^- b \bar{b}$ respectively. We would like to
stress that it is important to consider all the above topologies since
the expected numbers of events in the various channels never vanish
simultaneously for any value of $B$, as can be seen from expressions
(\ref{eq:nbb} -- \ref{eq:nllbb}). This allows us to obtain bounds on
$\epsilon_A^2$ and $\epsilon_B^2$ couplings without any assumptions on
the $h$ decay modes by varying $B$ from 0 to 1 and taking the weakest
limits.

In order to access the potentiality of LEP II to unravel de existence
of invisibly decaying Higgs bosons we assume that only the background
events were observed in accordance with the Tables~ \ref{tab:bbc} and
\ref{tab:ee}. Then, using Poisson statistics, we evaluate the
region of the five-dimensional parameter space ($M_h$, $M_A$,
$\epsilon_A$, $\epsilon_B$, $B$) that is excluded by this result at
95\% confidence level. Since this parameter space is quite large, we
make some simplifying assumptions below. For each channel, the
general form of the constraints on $\epsilon_A$ and $\epsilon_B$ is
\begin{equation}
c_A(M_h, M_A, B)~ \epsilon_A^2 ~+~ c_B(M_h, B)~ \epsilon_B^2 ~\le~
 n_0(M_h, M_A) \; ,
\end{equation}
where the functions $c_{A(B)}$ can be obtained from (\ref{eq:nbb}--
\ref{eq:nllbb}) and $n_0$ is the maximal allowed number of signal events, 
which depends on the background cross sections after cuts and on the
confidence level. It is clear from the above expression, that the
weakest limits on $\epsilon_A$ ($\epsilon_B$) can be obtained assuming
$\epsilon_B=0$ ($\epsilon_A=0$). In fact, for given values of $M_h$,
$M_A$, $B$, and $n_0$ the allowed region of the parameters
$\epsilon_A$ and $\epsilon_B$ is the interior of an ellipse with
semi-axes $\sqrt{n_0/c_A}$ and $\sqrt{n_0/c_B}$. Therefore, we present
the limits on the semi-axes of this ellipse as a function of $M_h$,
$M_A$ for the two most interesting cases: $B=1$, {\em i.e.} fully
invisible $h$ decay, and weakest limits, obtained by varying $B$ from 0 to 1.

For illustration, we exhibit in Table~ \ref{tab:maxn} typical values of
the 95\% CL maximum number of signal events in each channel ($n_0$),
assuming that the analysis is done for just one channel. These numbers
should be taken with a grain of salt since they depend on the point of
the parameter space due to the invariant mass cut.

\begin{table}[htbp]
\begin{center}
\begin{tabular}{|c|c|c|c|}
\hline
$\sqrt{s}$ (GeV) & ${\cal L}$ (pb$^{-1}$) & $b\bar{b}\protect\ptmis$
& $\ell^+\ell^-\protect\ptmis$ \\
\hline
175 & 500 & 4.70 & 6.36 \\
190 & 300 & 4.73 & 4.68 \\
205 & 300 & 6.33 & 5.93 \\
\hline
\end{tabular}
\caption{  \protect \small \baselineskip 12pt
Typical values of the 95\% CL maximum number of signal events in each
 channel, assuming that the analysis is done for just one channel.
}
\label{tab:maxn}
\end{center}
\end{table}

In order to obtain constraints on $\epsilon_A$ ($\epsilon_B$)
combining the different final state topologies, we calculate the
appropriate exclusion confidence levels CL$_i$ for each channel
separately, for a given value of $\epsilon_A$ ($\epsilon_B$). Then, we
evaluate the multi-channel exclusion confidence level using the
formulae
\begin{equation}
\mathrm{CL} = 1 - (1-\mathrm{CL}_1)...(1-\mathrm{CL}_n) \; .
\end{equation}
Finally, we choose as our limit for $\epsilon_A$ ($\epsilon_B$) the value
for which the combined CL is equal to 95\%.

\begin{figure}[htbp]
\begin{center}
\begin{tabular}{p{0.48\linewidth}p{0.48\linewidth}}
\begin{center}
\mbox{\epsfig{file=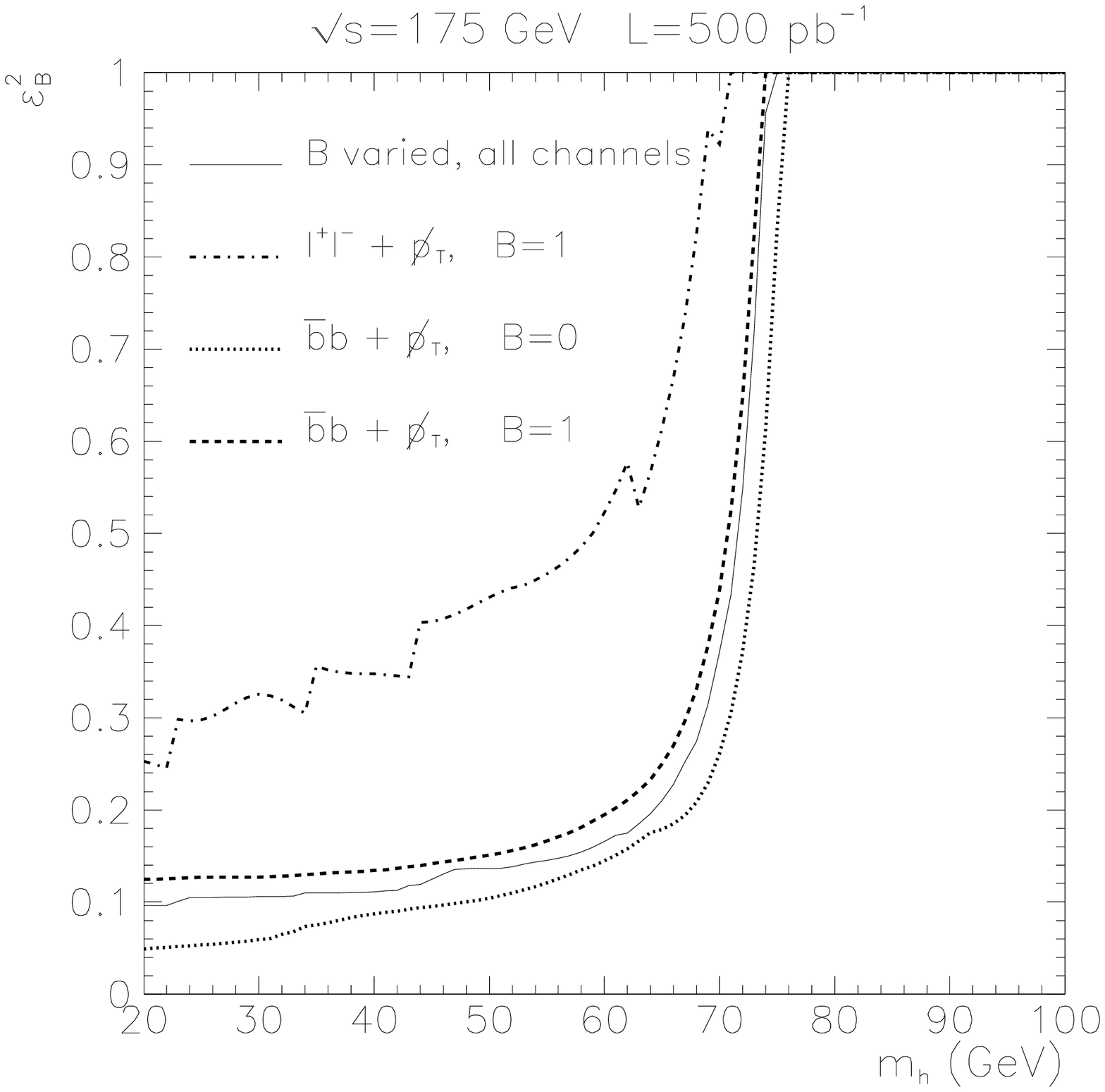,width=\linewidth,bbllx=10pt,bblly=10pt,%
bburx=530pt,bbury=530pt}}
\end{center}&
\begin{center}
\mbox{\epsfig{file=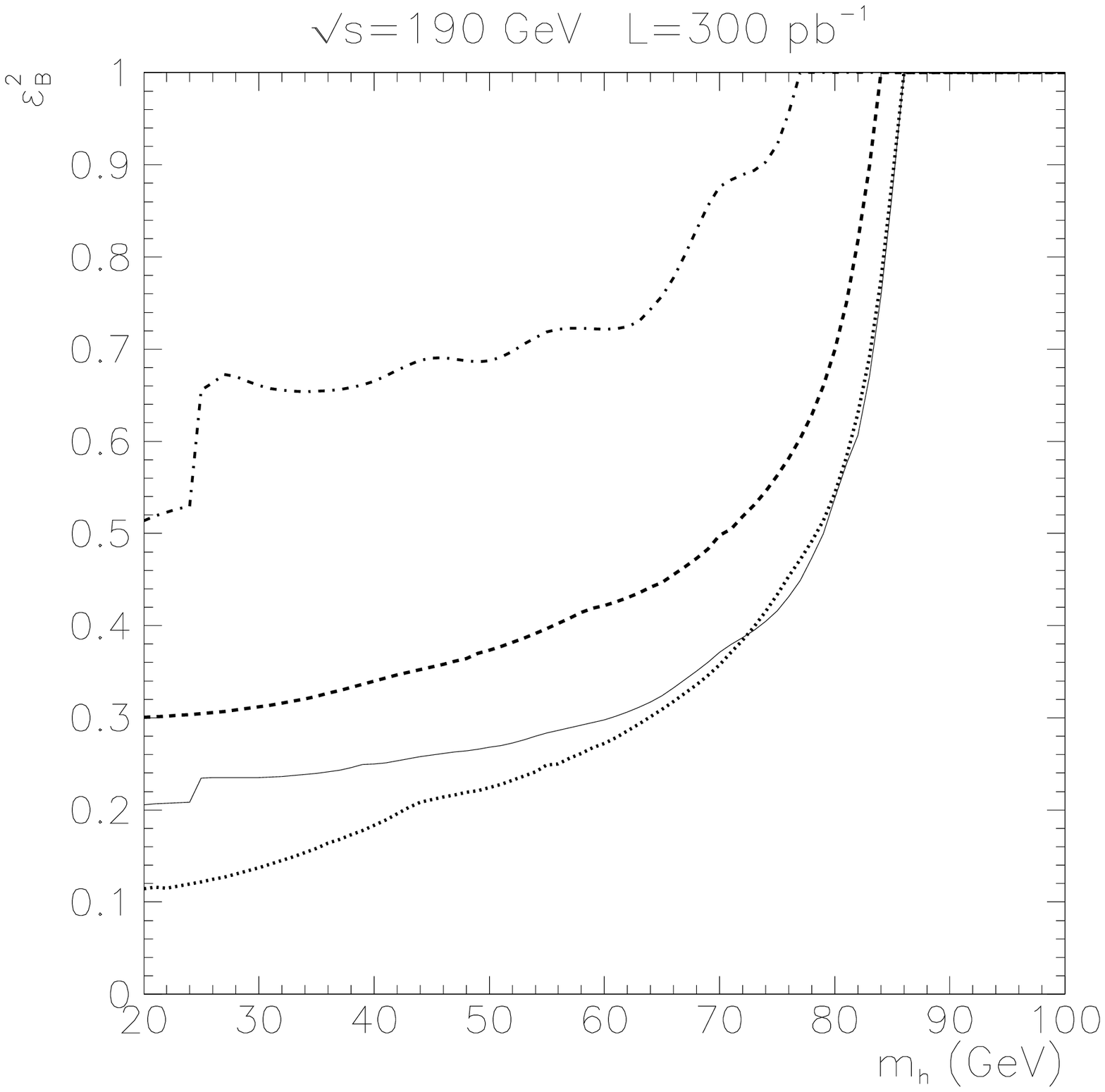,width=\linewidth,bbllx=10pt,bblly=10pt,%
bburx=530pt,bbury=530pt}}
\end{center}\\
\begin{center}
\mbox{\epsfig{file=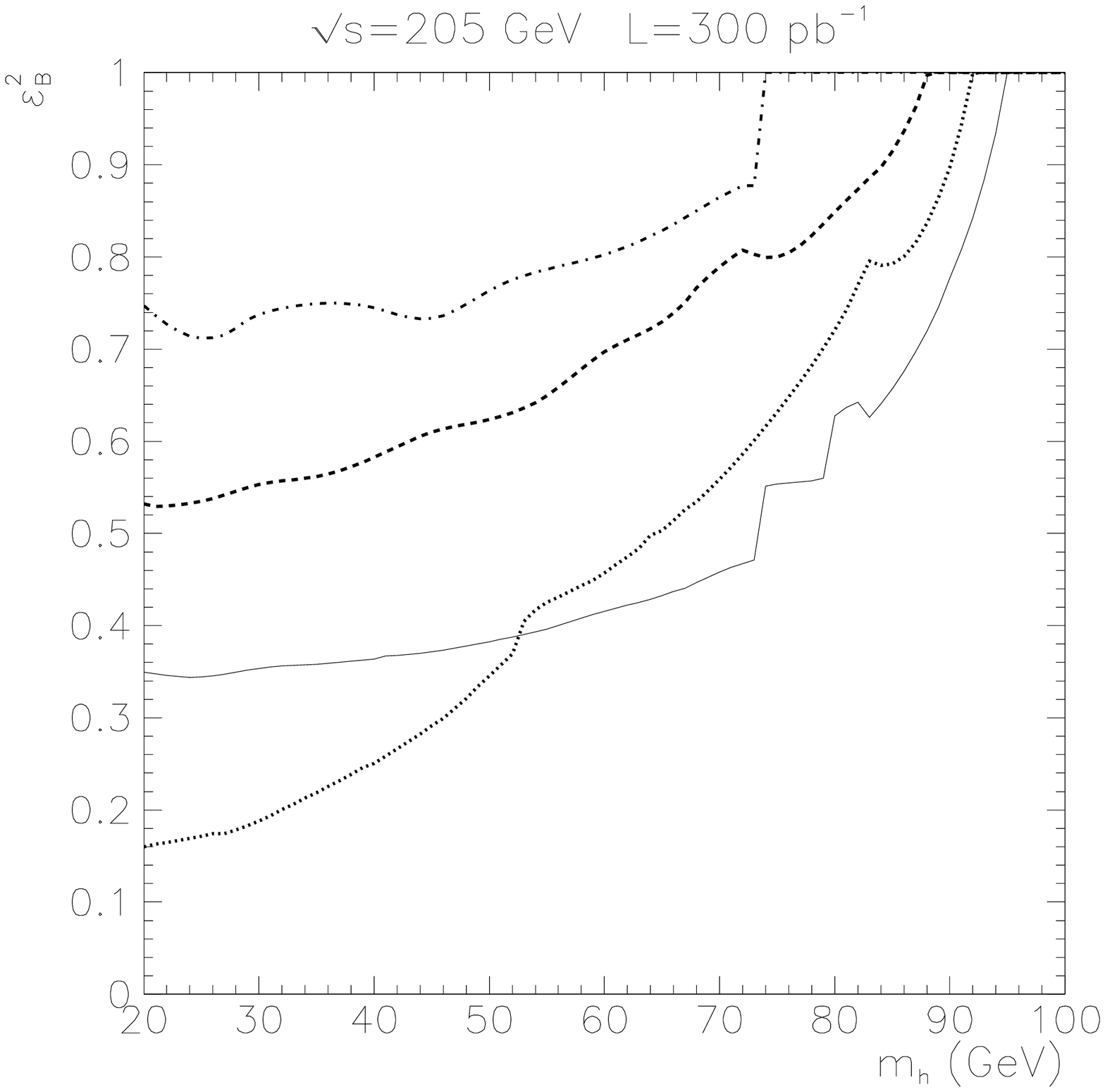,width=\linewidth,bbllx=10pt,bblly=10pt,%
bburx=530pt,bbury=530pt}}
\end{center}&
\begin{center}
\mbox{\epsfig{file=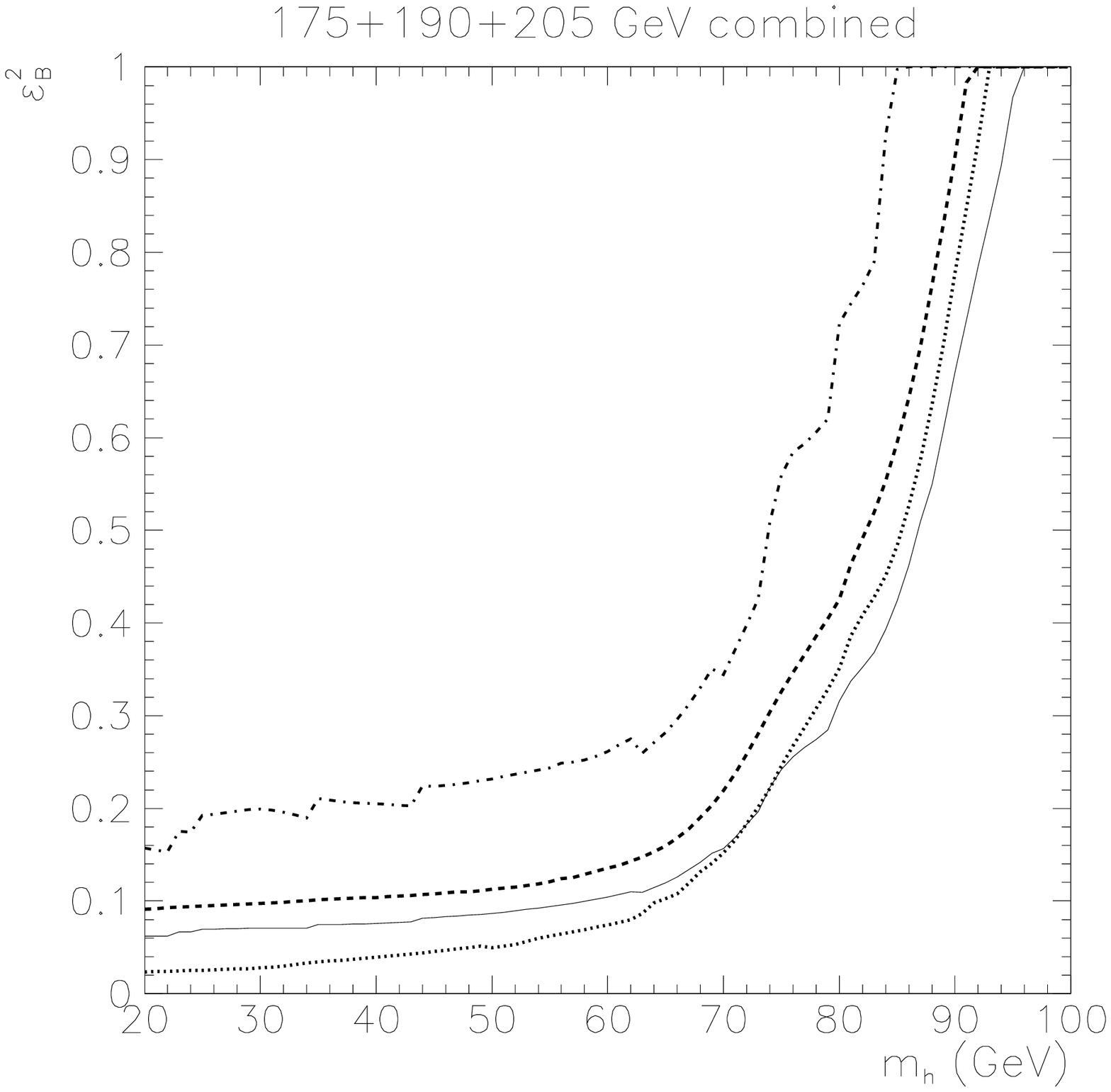,width=\linewidth,bbllx=10pt,bblly=10pt,%
bburx=530pt,bbury=530pt}}
\end{center}
\end{tabular}
\caption{\protect\small
\baselineskip 12pt
Bounds on $\epsilon_B^2$ as a function of $M_h$ for three values of
$\surd{s}$ and the three center-of-mass energies combined. See text
for further details.  }
\label{fig:zh}
\end{center}
\end{figure}

We start our analysis assuming that $\epsilon_A=0$ -- that is, we
study the simplest model exhibiting invisibly decaying Higgs bosons,
which is the one considered in Ref.\ \cite{valle:1,dproy}. In this
model only one singlet scalar field is added to the SM Higgs doublet.
We show in Fig.\ \ref{fig:zh} the constraints on the coupling
$\epsilon_B^2$, with the excluded region of the parameter space at
95\% CL being below the lines in this figure.  The dotted (dashed)
line stands for the constraints stemming from the $b\bar{b}\ptmis$
channel for $B=0$ ($B=1$), while the dot-dashed curve represents the
limits from the $\ell^+ \ell^- \ptmis$ channel for $B=1$.  We also
exhibit in this figure an absolute bound on $\epsilon_B^2$ (solid
line) based on all channels together, including the visible $h$ decays
(\ref{z:bb}) and (\ref{z:ll}).  The absolute bound is obtained by
varying $B$ in the range between $0$ and $1$ and taking the lowest
bound on $\epsilon_B^2$. The strongest single-channel constraints
originates from the $b\bar{b}\ptmis$ final state since there are many
more signal events with this topology, independently of the value of
the branching ratio $B$. Moreover, the $\ell^+ \ell^- \ptmis$ topology
also exhibits a relatively large $WW$ background. In fact, the
analysis of the final state $b\bar{b}\ptmis$ allow us to extend the
results of Ref.\ \cite{valle:1}.  Notice that for $B=0$ our limits are
in fact on the SM Higgs boson mass and on its coupling to the $Z$.
Indeed, our results are compatible to the ones obtained in Ref.\ 
\cite{DELPHI}.

\begin{figure}[htbp]
\begin{center}
\begin{tabular}{p{0.48\linewidth}p{0.48\linewidth}}
\begin{center}
\mbox{\epsfig{file=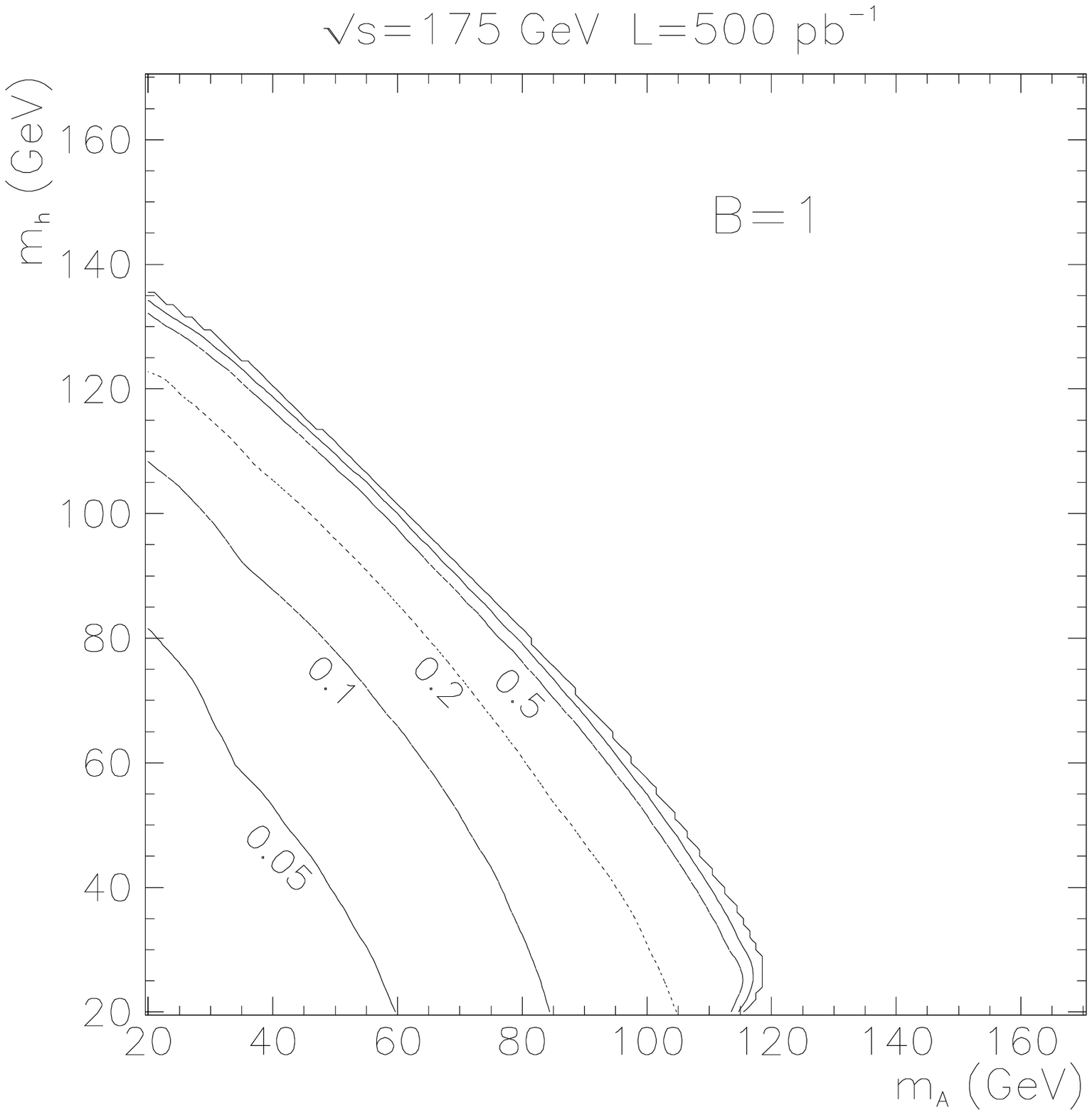,width=\linewidth,bbllx=10pt,bblly=10pt,%
bburx=530pt,bbury=530pt}}
\end{center}&
\begin{center}
\mbox{\epsfig{file=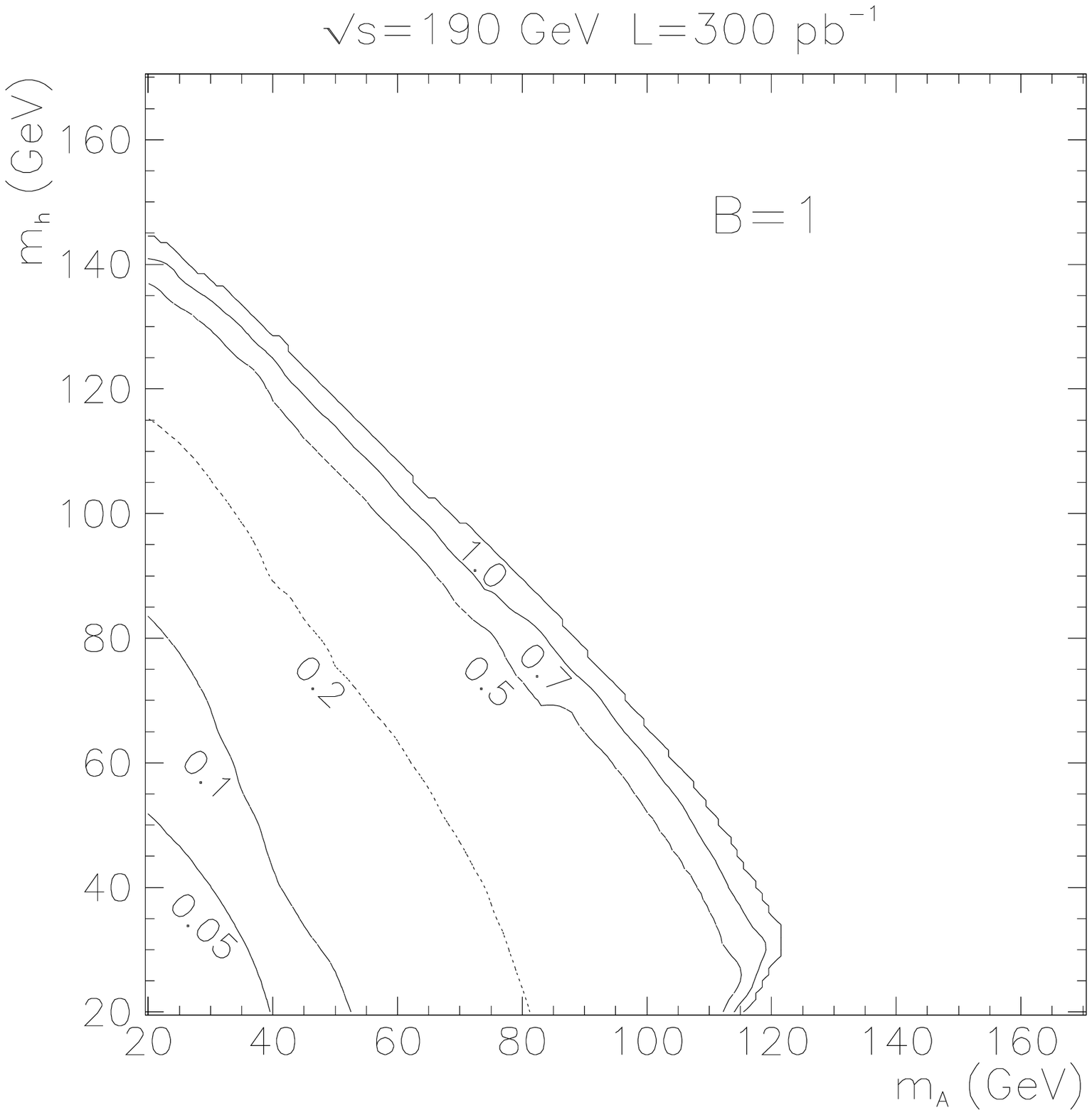,width=\linewidth,bbllx=10pt,bblly=10pt,%
bburx=530pt,bbury=530pt}}
\end{center}\\
\begin{center}
\mbox{\epsfig{file=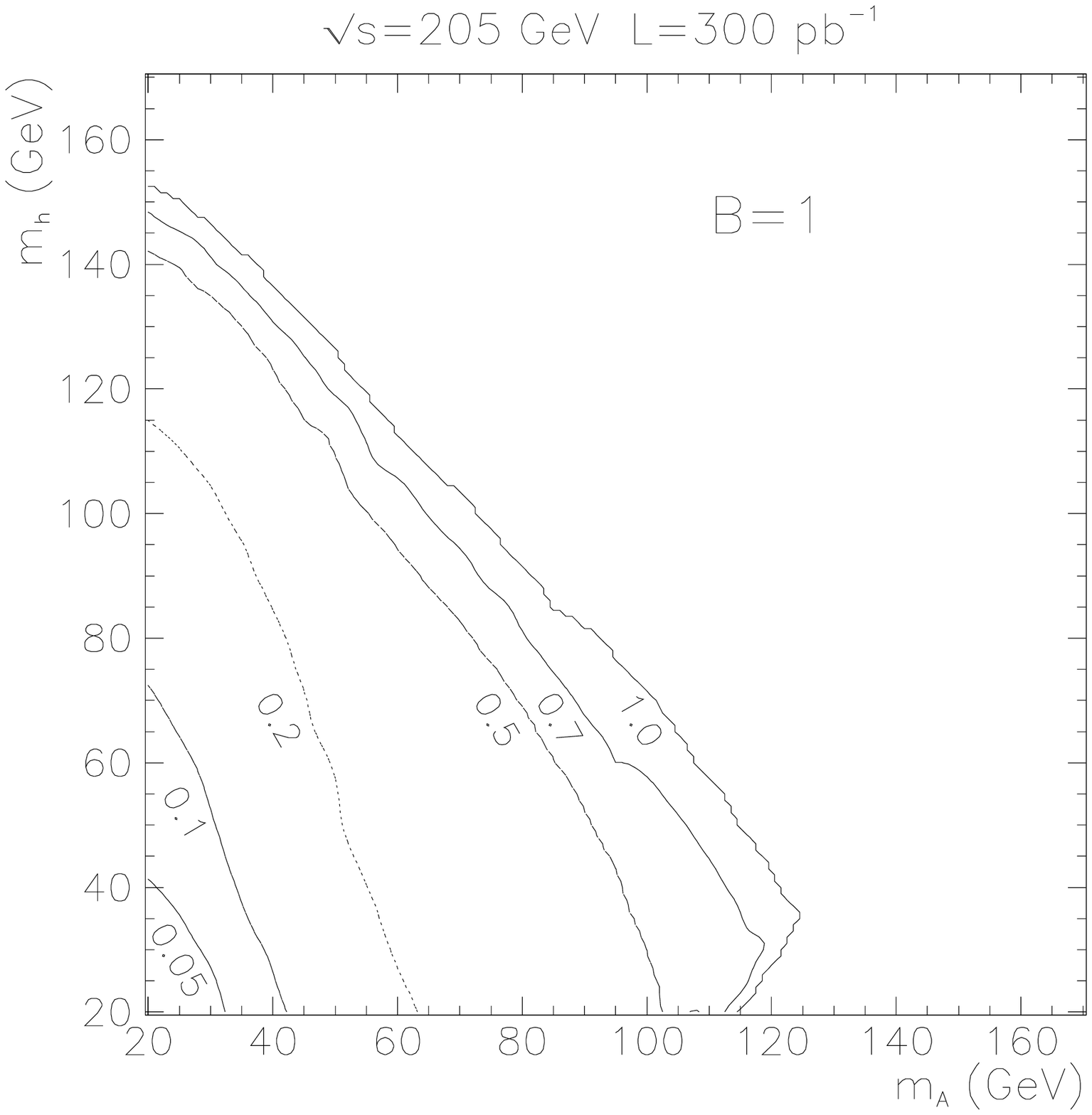,width=\linewidth,bbllx=10pt,bblly=10pt,%
bburx=530pt,bbury=530pt}}
\end{center}&
\begin{center}
\mbox{\epsfig{file=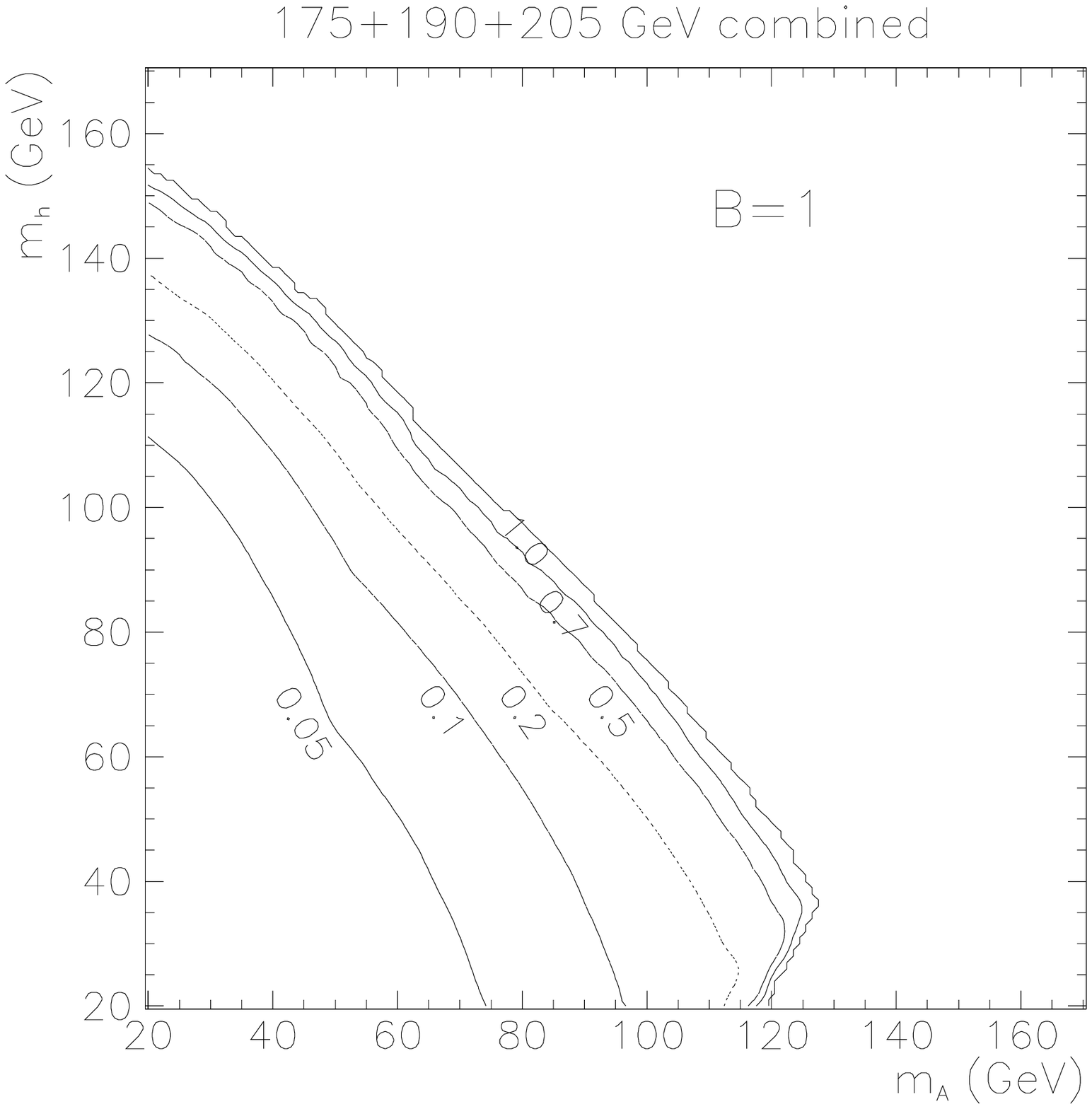,width=\linewidth,bbllx=10pt,bblly=10pt,%
bburx=530pt,bbury=530pt}}
\end{center}
\end{tabular}
\caption{\protect\small
\baselineskip 12pt
Bounds on $\epsilon_A^2$ as a function of $M_h$ and $ M_A$ for $B=1$.
The plots show the bounds obtained for $\surd{s} = 175$, $190$, and $205$
GeV and the constraints obtained by combining all three expected LEP
II runs. The allowed region of the parameter space is above the lines
of constant $\epsilon_A$.  }
\label{fig:iah}
\end{center}
\end{figure}

\begin{figure}[htbp]
\begin{center}
\begin{tabular}{p{0.48\linewidth}p{0.48\linewidth}}
\begin{center}
\mbox{\epsfig{file=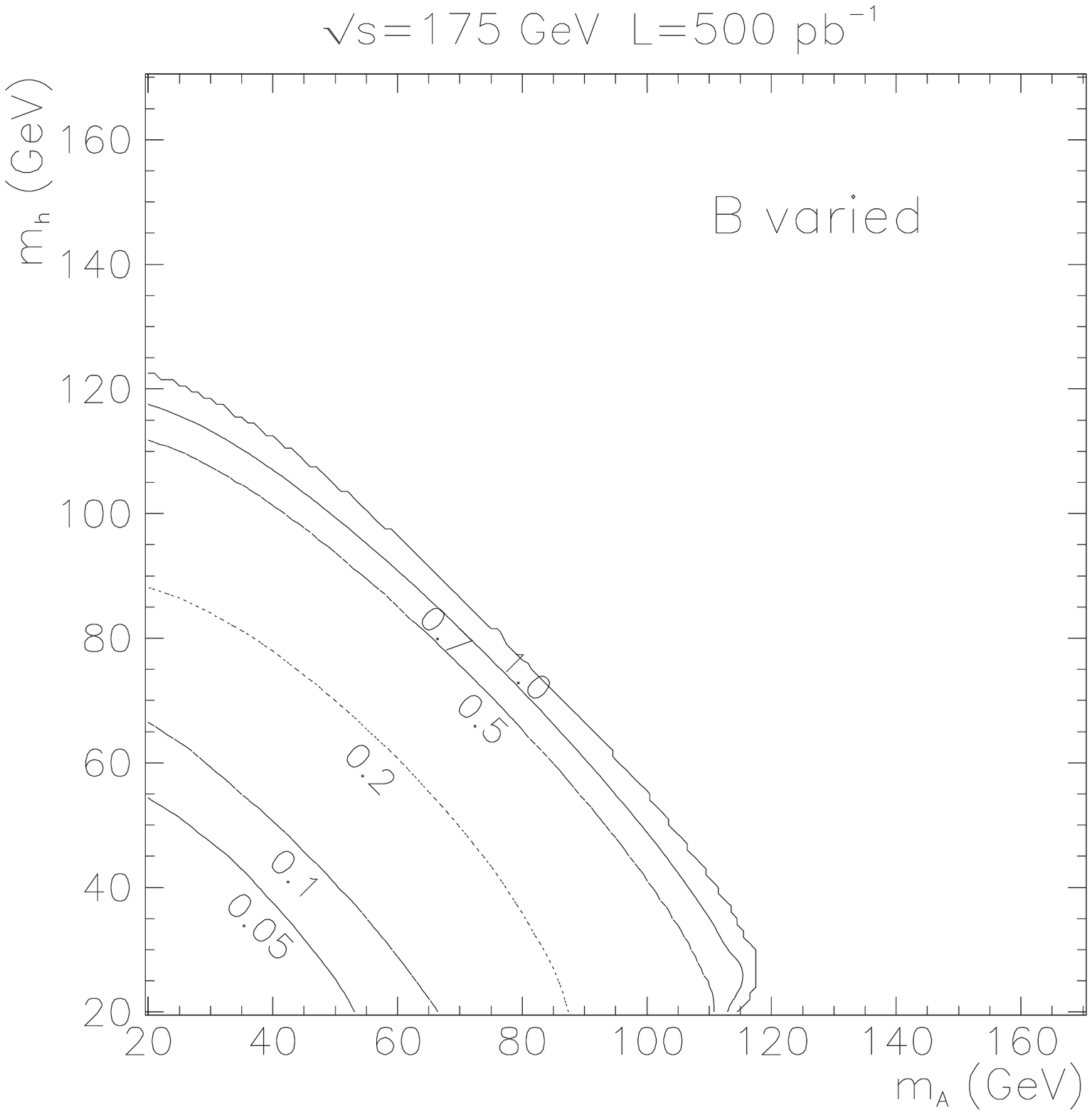,width=\linewidth,bbllx=10pt,bblly=10pt,%
bburx=530pt,bbury=530pt}}
\end{center}&
\begin{center}
\mbox{\epsfig{file=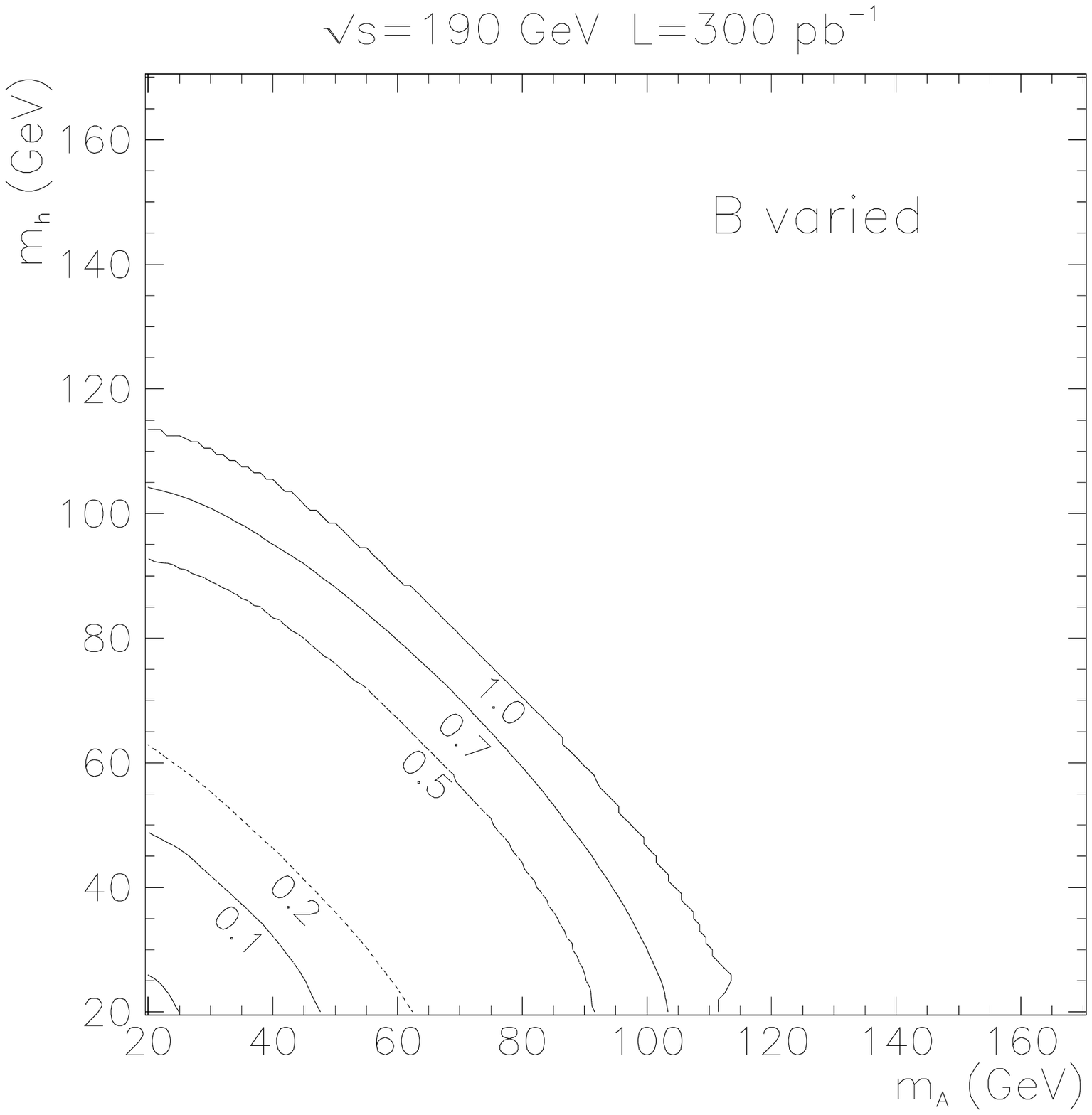,width=\linewidth,bbllx=10pt,bblly=10pt,%
bburx=530pt,bbury=530pt}}
\end{center}\\
\begin{center}
\mbox{\epsfig{file=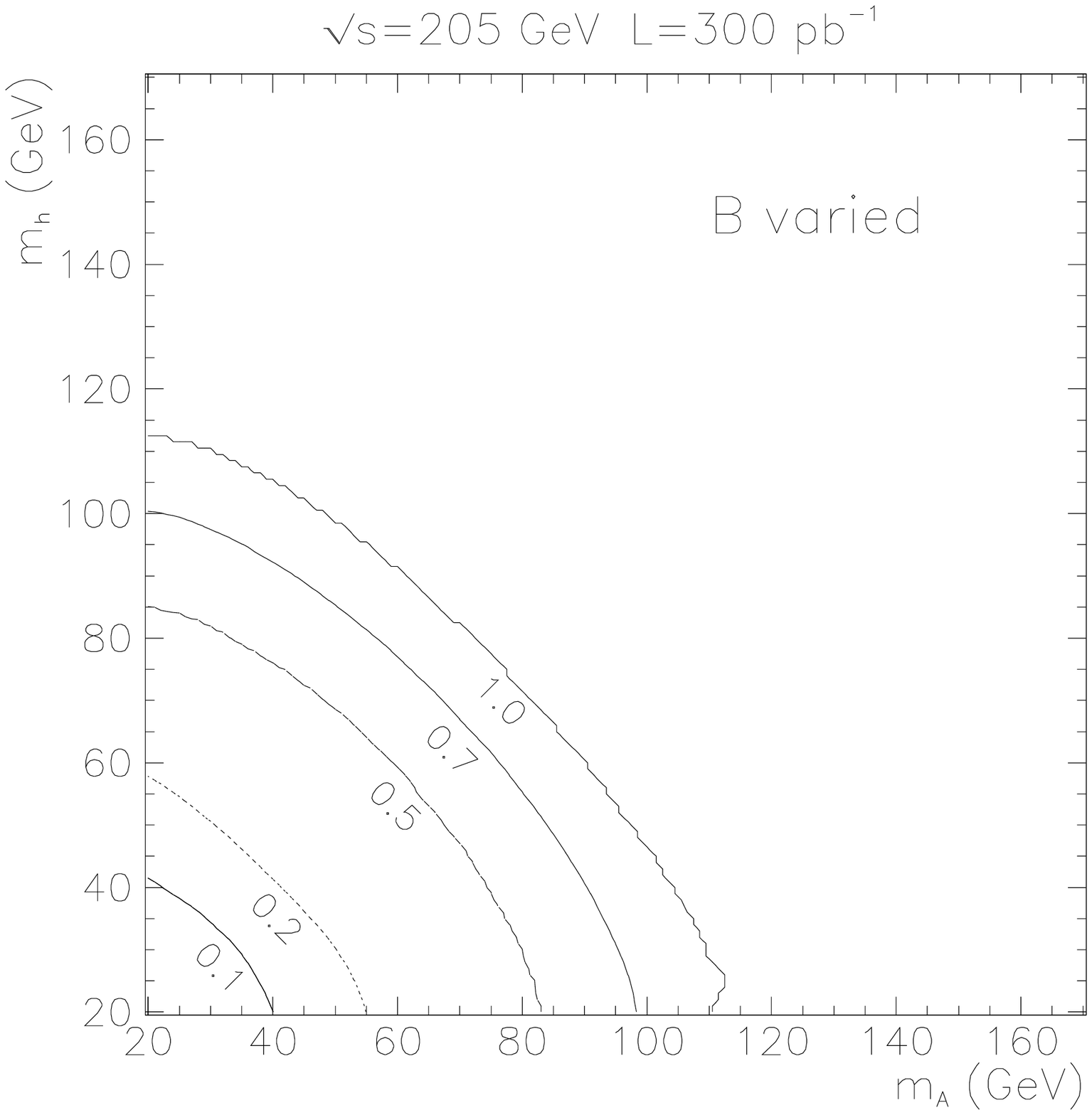,width=\linewidth,bbllx=10pt,bblly=10pt,%
bburx=530pt,bbury=530pt}}
\end{center}&
\begin{center}
\mbox{\epsfig{file=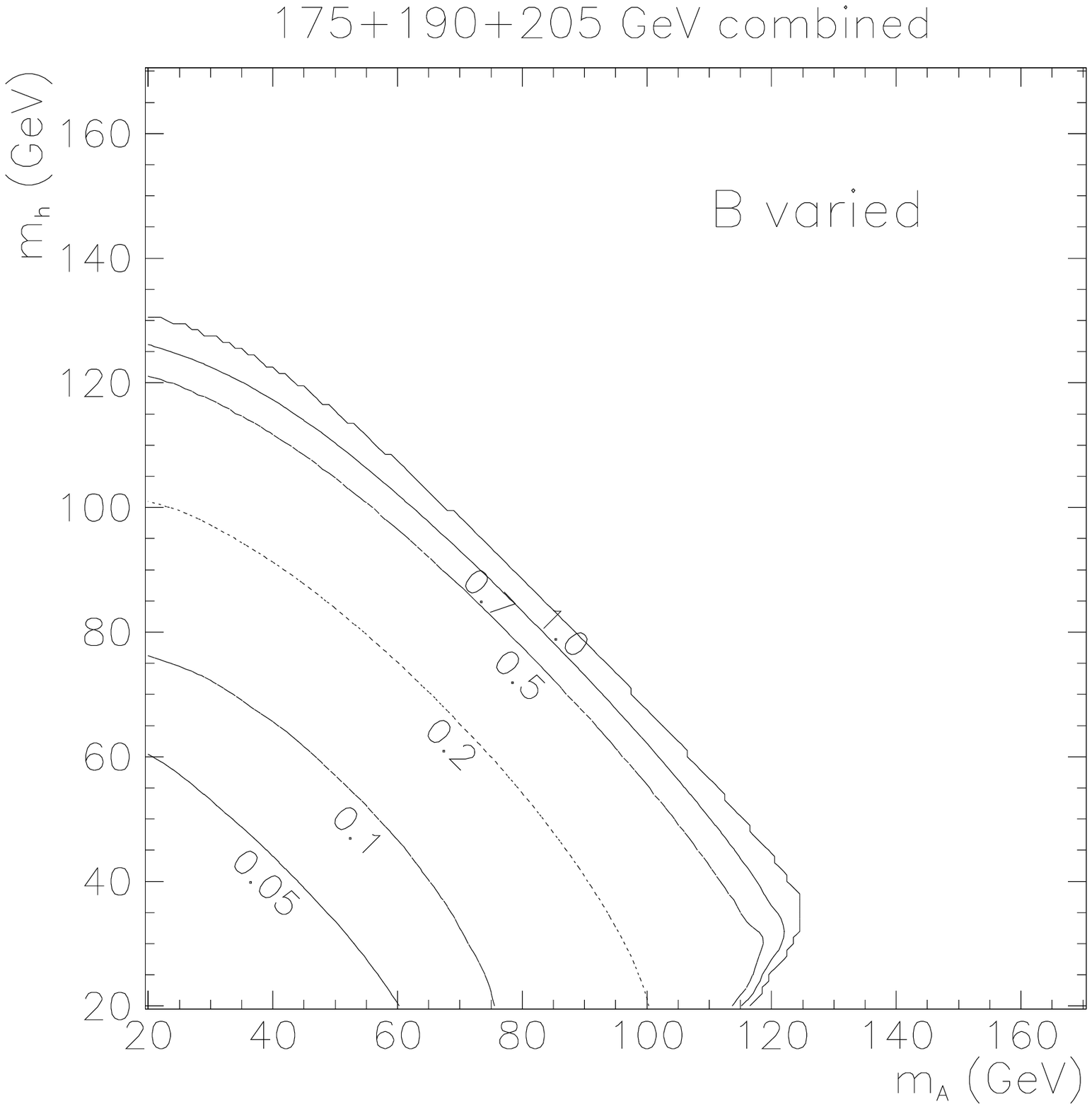,width=\linewidth,bbllx=10pt,bblly=10pt,%
bburx=530pt,bbury=530pt}}
\end{center}
\end{tabular}
\caption{\protect\small
\baselineskip 12pt
$B$-independent bounds on $\epsilon_A^2$ as a function of $M_h$ and 
$M_A$. The plots show the constraints obtained for $\surd{s} = 175$,
$190$, and $205$ GeV and the combined bounds from all three expected LEP
II runs.
}
\label{fig:ah}
\end{center}
\end{figure}

The weakest and more solid constraints on $\epsilon_A$ are those
obtained assuming $\epsilon_B=0$. Including all the final state
topologies and assuming $\epsilon_B=0$ we obtain the bounds on
$\epsilon_A^2$ which are shown in Figs.\ \ref{fig:iah} and
\ref{fig:ah}. Even with this simplifying hypothesis, we are still left
with the three-dimensional parameter space ($M_A$, $M_h$,
$\epsilon_A$). The absence of an invisible Higgs boson signal excludes
the region above the lines of constant $\epsilon_A$ in in Fig.\ 
\ref{fig:iah}, at 95\% CL.  In this figure we exhibit the constraints
on $\epsilon_A^2$ obtained under the assumption that $h$ decays
exclusively into the invisible final state ($B=1$).  Fig.\ 
\ref{fig:ah} contains the $B$-independent bounds on $\epsilon_A^2$
obtained, as in the $\epsilon_B^2$ case, by varying $B$ in the range
between $0$ and $1$ and keeping the weakest bound.

In general, the topologies (\ref{eq:nbb}) and (\ref{eq:n4b}) are
dominated by the associated production, as long as they are not
strongly suppressed by small $\epsilon_A$ couplings or by phase space.
Therefore, for a given value of $M_h$, the constraints on the
associated production coupling $\epsilon_A$ are stronger than those on
$\epsilon_B^2$ provided $M_A$ is not very large. Another general
feature of our results is that the final state $b\bar{b}
\ptmis$ (\ref{eq:nbb}) leads to the stronger limits on $\epsilon_A$
coupling than those given by the other topologies where $h$ is
decaying visibly, especially by $b\bar{b}b\bar{b}$ final state.

%^^^^^^^^^^^^^^^^^^^^^^^^^^^^^^^^^^^^^^^^^^^^^^^^^^^^^^^^^^^^^^^^^^^^^

\section{Conclusions}

The Higgs boson can decay into a pair of invisible massless Goldstone
bosons in a wide class of models in which a global symmetry, such as
lepton number, is spontaneously broken. We performed a
model-independent analysis of the capability of LEP II to probe for
such a Higgs boson, assuming that it couples to the $Z$ and a CP-odd
scalar $A$, not analysed before. We studied the final state topologies
$b \bar{b} \ptmis$, $\ell^+ \ell^- \ptmis$, $\ell^+ \ell^- b \bar{b}$,
and $b \bar{b} b \bar{b}$, taking into account the backgrounds and
choosing the cuts so as to enhance the signal.

In the case that the invisible Higgs boson does not couple to the
CP-odd scalar $A$ ($\epsilon_A=0$), we found out that the strongest
constraints on the parameter space come from the final state $b
\bar{b} \ptmis$. For masses $M_h$ up to approximately $70$ GeV, the
planned run at $\sqrt{s} = 175$ leads to the strongest limits due to
its higher luminosity. On the other hand, the higher centre-of-mass
energy runs are needed to expand the range of masses that can be
probed. The results of our complete analysis for this case extend the
previous results of Ref.\ \cite{valle:1}. We also analysed the extreme
case $\epsilon_B=0$, in which the invisibly decaying Higgs boson does
not couple to the $Z$.  In this scenario, the strongest limits also
come from the final state $b\bar{b} \ptmis$. As a rule of thumb, the
signal for the invisible Higgs being produced in association with $A$
can detected provided $M_h + M_A \lsim 150 $ (100) GeV for $\epsilon_A
= 1$ (0.1).

The invisibly decaying Higgs boson can also give rise to signals at
the LHC, such as $\ell^+ \ell^- \ptmis$ \cite{lhc}.  The invisible
decay has good advantages over the standard model $h \rightarrow
\gamma\gamma$ decay mode in the intermediate Higgs mass region, since
its branching fraction can be large. Unfortunately, however, the
ability to reconstruct the invisible Higgs boson mass is absent in the
case of hadron collisions. This makes the signature of invisibly
decaying Higgs bosons in $e^+ e^-$ collisions especially important and
a crucial check of any signal that might be seen at the LHC.  In this
paper we have shown that LEP II will be able to unravel the existence
an invisibly decaying Higgs boson for a large fraction of the relevant
parameter space. As a final remark we note that models with invisibly
decaying Higgs bosons may lead to other interesting physical effects
that could be detectable experimentally \cite{granada}.

%===============================================================

\vskip 1.0 cm

\begin{center}
{\bf ACKNOWLEDGMENTS}
\end{center}

We thank S. Katsanevas for useful discussions and for bringing 
reference \cite{DELPHI} to our attention. This work was supported by
the University of Wisconsin Research Committee with funds granted by
the Wisconsin Alumni Research Foundation, by the U.S.\ Department of
Energy under contract No.  DE-FG02-95ER40896, by DGICYT under grant
No. PB92-0084, by Conselho Nacional de Desenvolvimento Cient\'{\i}fico
e Tecnol\'ogico (CNPq/Brazil), by Fundac\~ao de Amparo \`a
Pesquisa do Estado de S\~ao Paulo (FAPESP/Brazil), and by a DGICYT
postdoctoral fellowship and the Alexander von Humboldt Stiftung.

%^^^^^^^^^^^^^^^^^^^^^^^^^^^^^^^^^^^^^^^^^^^^^^^^^^^^^^^^^^^^^^^^^^^^^^

\newpage

%%%%%%%%%%%%%%%%

%%%%%%%%%%%%%%%%%%%%

%%%%%%%%%%%%%%%%%%%

%%%%%%%%%%%%%%%%%%%%%%%%%%%%%%%%%%%%%%%%%

\end{document}